\pdfoutput=1

\documentclass[11pt,twoside,a4paper,cmspaper,final,collab]{cms-tdr}
\def\svnVersion{420604P}\def\cmsCernNoTag{CERN-EP-2017-090}\def\cmsCernDate{\today}\def\cmsMessage{Published in the Journal of High Energy Physics as \href{http://dx.doi.org/10.1007/JHEP08(2017)029}{\doi{10.1007/JHEP08(2017)029.}}}

\begin{document}\cmsNoteHeader{B2G-16-016}

\hyphenation{had-ron-i-za-tion}
\hyphenation{cal-or-i-me-ter}
\hyphenation{de-vices}
\RCS$Revision: 416195 $
\RCS$HeadURL: svn+ssh://svn.cern.ch/reps/tdr2/papers/B2G-16-016/trunk/B2G-16-016.tex $
\RCS$Id: B2G-16-016.tex 416195 2017-07-13 22:08:40Z drankin $
\newcommand{\mtb}{\ensuremath{M_{\PQt\PQb}}\xspace}
\providecommand{\PWRpr}{\ensuremath{\PW'_\mathrm{R}}\xspace}
\newcommand{\mc}[3]{\multicolumn{#1}{#2}{#3}}
\cmsNoteHeader{B2G-16-016}
\title{Searches for $\PWpr$ bosons decaying to a top quark and a bottom quark in proton-proton collisions at 13\TeV}

\date{\today}

\abstract{
Searches are presented for heavy gauge bosons decaying into a top and a bottom quark in data collected by the CMS experiment at $\sqrt{s} = 13\TeV$
that correspond to an integrated luminosity of 2.2 and 2.6\fbinv in the leptonic and hadronic analyses, respectively.
Two final states are analyzed, one containing a single electron, or muon, and missing transverse momentum, and the other containing multiple jets and no electrons or muons.
No evidence is found for a right-handed $\PWpr$ boson ($\PWRpr$) and the combined analyses exclude
at 95\% confidence level $\PWRpr$ with masses below 2.4\TeV if $M_{\PWRpr}\gg M_{\nu_\mathrm{R}}$ (mass of the right handed neutrino), and below 2.6\TeV if $M_{\PWRpr}<M_{\nu_\mathrm{R}}$.
The results provide the most stringent limits for right-handed $\PWpr$ bosons in the top and bottom quark decay channel.
}

\hypersetup{%
pdfauthor={CMS Collaboration},%
pdftitle={Searches for W' bosons decaying to a top quark and a bottom quark in proton-proton collisions at 13 TeV},%
pdfsubject={CMS},%
pdfkeywords={CMS, physics, B2G, exotica, Wprime}
}

\maketitle

\section{Introduction}

Many theories that extend the standard model (SM) predict additional charged gauge bosons~\cite{doi:10.1146/annurev.nucl.55.090704.151502,PhysRevD.64.035002,PhysRevD.64.065007,PhysRevD.53.5258,PhysRevD.11.566},
often referred to  as $\PWpr$ bosons.
In models where the resonance is sufficiently massive, it is common to postulate that the coupling to third generation quarks might be enhanced relative to the second and first generations~\cite{Muller1996345,Malkawi1996304}, making a search for the decay $\PWpr\to\PQt\PAQb$ or $\PAQt\PQb$ highly appropriate.
A particular advantage of this kind of search is that this channel is more easily distinguished from the large continuum of multijet background than searches in the decays to light quarks ($\PWpr\to \qqbar$).
The search in top and bottom quark (tb) systems complements searches in $\PWpr\to\ell\nu$ (where $\ell$ denotes a charged lepton and $\nu$ denotes a neutrino) and $\PWpr\to \mathrm{VV}$ (where $\mathrm{V}$ denotes an SM $\PW$ or \Z boson) channels.
The tb final state also benefits from the fact that its $\PWpr$ mass can be fully determined, whereas in the leptonic mode there is a two-fold ambiguity in its mass.

This paper presents the first search performed for a right-handed $\PWpr$ ($\PWRpr$) decaying to a top and a bottom quark at $\sqrt{s}=13\TeV$,
using data collected by the CMS experiment corresponding to an integrated luminosity of up to 2.6\fbinv.
In scenarios where a theoretical right-handed neutrino ($\nu_R$) is heavier than the $\PWRpr$, the decay $\PWRpr\to\ell\nu_\mathrm{R}$ is forbidden and the branching fraction $\mathcal{B}(\PWRpr\to\PQt\PQb)$ is enhanced.
This makes the $\PWRpr\to\PQt\PQb$ decay an important channel in the search for $\PWpr$ bosons.
Previous searches in the tb channel have been performed at the Fermilab Tevatron~\cite{D0:2010,D0Wprime,PhysRevLett.115.061801} and at the CERN LHC by both the CMS~\cite{Chatrchyan:2014koa,Chatrchyan:2016had} and ATLAS~\cite{Aad:2014xea, Aad:2014xra} Collaborations.
The most stringent limits to date on the production of $\PWpr$ bosons with purely right-handed couplings come from the CMS search performed at $\sqrt{s}=8\TeV$~\cite{Chatrchyan:2016had}.
Relative to this 8\TeV search, the expected production cross section of the $\PWRpr$ boson at $\sqrt{s}=13$\TeV is enhanced by a factor of approximately 7\,(13) for a 2 (3)\TeV resonance.

We separately analyze events with and without a lepton in the final state (referred to as leptonic and hadronic analyses), and then combine the results.
In both analyses, the invariant mass of the tb system ($\mtb$) is used to conduct searches for the $\PWRpr$ boson.
The achieved sensitivity after combining the results is better than in each individual channel, thereby providing improved exclusion limits compared to previous results.

In this paper, Section~\ref{sec:detect} contains a description of the CMS detector.
Section~\ref{sec:model} provides details of the simulated samples and their production, while Section~\ref{sec:rec} discusses the techniques used for object reconstruction and event selection.
The methods used for estimation of backgrounds are given in Section~\ref{sec:anstrat}.
Section~\ref{sec:syst} provides information on systematic uncertainties, and Section~\ref{sec:results} presents results of the individual and combined analyses.
A summary is given in Section~\ref{sec:summary}.
\section{The CMS detector\label{sec:detect}}

The central feature of the CMS apparatus is a superconducting solenoid of 6\unit{m} internal diameter, providing a magnetic field of 3.8\unit{T}.
A silicon pixel and strip tracker, a lead tungstate crystal electromagnetic calorimeter (ECAL), and a brass and scintillator hadron calorimeter (HCAL), each composed of a barrel and two endcap sections, reside within the solenoid field.
Forward calorimeters extend the pseudorapidity ($\eta$) coverage~\cite{Chatrchyan:2008zzk} provided by the barrel and endcap detectors.
Muons are measured in gas-ionization detectors embedded in the steel flux-return yoke outside the solenoid.

The electron momentum is estimated by combining the energy measurement in the ECAL with the momentum measurement in the tracker.
The momentum resolution for electrons with $\pt\approx45\GeV$ from $\Z \to \Pep \Pem$ decays ranges from 1.7\% for electrons without an accompanying shower in the barrel region, to 4.5\% for electrons showering in the endcaps~\cite{Khachatryan:2015hwa}.

Muons are measured in the range $\abs{\eta} < 2.4$, with detection planes based on drift tubes, cathode strip chambers, and resistive plate chambers.
Matching muons to tracks in the silicon tracker yields a relative $\pt$ resolution for muons with $20<\pt<100\GeV$ of 1.3--2.0\% in the barrel and better than 6\% in the endcaps.
The \pt resolution in the barrel is better than 10\% for muons with \pt up to 1\TeV~\cite{Chatrchyan:2012xi}.

Events of interest are selected using a two-tiered trigger system~\cite{Khachatryan:2016bia}.
The first level (L1), composed of custom hardware processors, uses information from the calorimeters and muon detectors to select events
at a rate of around 100\unit{kHz} within a time interval of less than 4\mus.
The second level, known as the high-level trigger (HLT), consists of a farm of processors running a version
of the full event reconstruction software optimized for fast processing,
and reduces the event rate to less than 1\unit{kHz} before data storage.

The particle-flow event algorithm~\cite{CMS-PAS-PFT-09-001,CMS-PAS-PFT-10-001,Sirunyan:2017ulk} reconstructs and identifies each individual particle candidate using an optimized combination of information from the various elements of the CMS detector.
The energy of photons is obtained from the ECAL measurement, and corrected for the online suppression of signals close to threshold.
The energy of electrons is determined from a combination of the electron momentum at the primary interaction vertex determined by the tracker, the energy of the corresponding ECAL cluster, and the energy sum of all bremsstrahlung photons spatially compatible with originating from the electron track.
The energy of muons is obtained from the curvature of the corresponding track. The energy of charged hadrons is determined from a combination of their momentum measured in the tracker and the matching ECAL and HCAL energy deposition, corrected for suppression of small signals and for the response of hadron showers in the calorimeters.
Finally, the energy of neutral hadrons is obtained from the corresponding corrected ECAL and HCAL energies.

The missing transverse momentum vector, \ptvecmiss, is defined as the projection on the plane perpendicular to the beams of the negative vector sum of the momenta of all reconstructed particles in an event.

A more detailed description of the CMS detector, together with a definition of the coordinate system and the kinematic variables, can be found in Ref.~\cite{Chatrchyan:2008zzk}.
\section{Modeling of signal and background \label{sec:model}}

All signal events are generated at leading order (LO) using the CompHEP 4.5.2 \cite{comphep} package and their cross sections are scaled to next-to-leading order (NLO)
with an approximate K-factor of 1.2 \cite{Sullivan:2002jt,Duffty:2012rf}.
All signal samples are generated with purely right-handed couplings, according to the following model-independent, lowest-order, effective Lagrangian:
\begin{equation}
 \mathcal{L}=\frac{V_{f_if_j}}{2\sqrt{2}}g_\PW\bar{f}_i\gamma_\mu(1+\gamma^5)\textrm{W}'^\mu f_j+\textrm{H.C.},
\end{equation}
where $V_{f_if_j}$ is the element of the Cabibbo--Kobayashi--Maskawa matrix when $f$ is a quark, and $V_{f_if_j}=\delta_{ij}$ when $f$ is a lepton, and $g_\PW$ is the SM weak coupling constant.
Since we consider $\PWRpr$ bosons (with right-handed couplings), there is no interference at production with the SM W boson.
The simulation for leptonic decays of the $\PWRpr$ boson includes decays involving a $\tau$ lepton,
and no distinction is made in the analysis between an electron or muon directly from the W boson decay and an electron or muon from a subsequent leptonic $\tau$ decay.
Signal samples are generated for signal masses between 1 and 3\TeV in 100\GeV steps.
The width of the $\PWRpr$ generated by {\sc CompHEP} is narrow, and varies with the mass, but is approximately 3\% for all masses considered in this analysis.
This is smaller than the invariant mass resolution of the detector, and therefore the precise values of the width does not have a significant effect on our results.

For right-handed \PWpr bosons, the leptonic decays necessarily produce right-handed neutrinos ($\nu_\mathrm{R}$).
When the mass of the $\nu_\mathrm{R}$
is larger than that of the $\PWRpr$ boson ($M_{\PWRpr}<M_{\nu_\mathrm{R}}$) then the $\PWRpr\to\ell\nu_\mathrm{R}$ decays are kinematically forbidden and only $\PWRpr\to\textrm{q}\bar{\textrm{q}}'$ decays are allowed (of which $\PWRpr\to\PQt\PQb$ is a subset).
On the other hand, if the $\nu_\mathrm{R}$ is lighter than the $\PWRpr$ boson ($M_{\PWRpr}>M_{\nu_\mathrm{R}}$) then $\PWRpr\to\ell\nu_\mathrm{R}$ decays are allowed.
Consequently, the product of the $\PWRpr$ cross section and its branching fraction ($\PWpr \to $ tb) is enhanced for heavy neutrinos by approximately one third.
When calculating the distribution in the number of expected signal events, it is always assumed
that $M_{\PWRpr}\gg M_{\nu_\mathrm{R}}$.
When displaying upper limits at 95\% confidence levels (CL), we consider both scenarios.

The SM processes that contribute significantly to the background in the leptonic analysis are W+jets and \ttbar events.
The background in the hadronic analysis is dominated by multijet and \ttbar production.
Although it is a much smaller contribution to the total background, both analyses also consider associated production of a top quark and a W boson as background, while the leptonic analysis further considers both $t$- and $s$-channel single top quark, Z or $\gamma^*$+jets, and diboson (WW, WZ and ZZ) production.
The hadronic and leptonic analyses employ different methods of background estimation because of differences in the final states.
All background predictions from nondominant sources are estimated from simulation.

Simulated samples for $\Z/\gamma^*$+jets, $s$-, and $t$-channel single top, and W+jets are generated at NLO using the
{\MGvATNLO}~\cite{Alwall:2014hca,Frederix:2012ps,Alwall:2007fs} v2.2.2 generator. The $\ttbar$ and single top quark in the tW channel samples are generated using the {\POWHEG}
v2 generator~\cite{Nason:2004rx,powheg,Alioli:2010xd,Frixione:2007nw,Re:2010bp}, and all other backgrounds are generated at LO using the \PYTHIA 8.2~\cite{Sjostrand:2014zea} generator.
In all cases, NNPDF 3.0 parton distribution functions (PDFs) are used~\cite{nnpdf}.

Both hadronic and leptonic analyses use the MC simulated \ttbar background prediction.
In the leptonic analysis, the \ttbar simulation is assigned a correction based on the top quark $\pt$, which is known to be improperly modeled~\cite{Khachatryan:2015oqa}.
This correction is not necessary in the hadronic analysis because of differences in the phase space resulting from the specific event selections, and is confirmed in a $\ttbar$ enriched control region.
The predictions from both analyses are checked in control regions that are independent with respect to the signal region and contain minimal contamination from signal.
In both cases, the agreement between the data and prediction from simulation is good.

For the W+jets background in the leptonic analysis, the initial prediction is estimated from simulation.
The agreement with data is then checked in a control region dominated by W+jets events.
The same region is also used to extract correction factors for different W+jets components, e.g., W+light-quark or gluon jets and W+charm or bottom quark jets.
The relative composition of these components in simulation is known to differ~\cite{Chatrchyan:2011yy} from the composition in data, and we apply these correction factors to the predictions.

The multijet background in the hadronic analysis is determined from data in independent control regions. The validity of the estimation procedure is then checked using simulated multijet events.

More details on the background estimation methods can be found in Section \ref{sec:anstrat}.

All simulated signal and background events are processed through \PYTHIA 8.2 for parton fragmentation and hadronization, where the underlying
event tune CUETP8M1~\cite{Khachatryan:2015pea} has been used.
The simulation of the CMS detector is performed using \GEANTfour~\cite{geant}.
Also, all simulated event samples include additional overlapping proton-proton interactions in the same or adjacent bunch crossings (pileup)
that are weighted such that the distribution in the number of interactions agrees with that expected in data.
\section{Event reconstruction and selection\label{sec:rec}}

The two analyses employ
different selections targeted at their respective signal topologies.
Details on specific aspects of the selections are given below.

\subsection{Jet reconstruction}

Jets are reconstructed offline from the particle-flow candidates, clustered using the anti-\kt algorithm~\cite{Cacciari:2008gp, Cacciari:2011ma} with distance parameters of 0.4 (AK4 jets) and 0.8 (AK8 jets).

The jet momentum is defined by the vectorial sum of all particle-flow candidate momenta in the jet, and is found from simulation to be within 5 to 10\% of the true momentum.
An offset correction is applied to jet momenta to take into account the contribution from pileup.
Jet energy corrections~\cite{jec} are obtained from simulation, and are confirmed with in situ measurements of the energy balance in dijet and photon+jet events.
Additional selection criteria are applied to each event to remove spurious jet-like features originating from isolated noise patterns in certain HCAL regions.

Both the leptonic and hadronic analyses use the charged-hadron subtraction method, which removes from the event any charged hadrons not associated with the leading vertex, defined as the vertex with the highest $\pt^{2}$ sum.
The estimated contribution from pileup to the neutral hadron component of jets is also subtracted, based on the jet area~\cite{areasubtract}.

The leptonic analysis uses AK4 jets because their smaller area makes them less sensitive to pileup, and
the hadronic analysis uses AK8 jets whose larger area makes them more suited to the jet substructure-based techniques used to identify highly Lorentz-boosted top quark decays.
These techniques are discussed in Section~\ref{sec:ttag}.

\subsubsection{Identification of b jets}

The combined secondary vertex version 2 (CSVv2) algorithm~\cite{btagging,CMS-PAS-BTV-15-001}, which combines secondary vertex and track based lifetime information in order to identify b jets, is used by both analyses.
They use an operating point which has a b jet identification (b tagging) efficiency of 80\% and a light-flavor jet misidentification (mistag) probability of 10\%.
A scale factor is applied as a function of \pt to correct observed differences in performance between data and simulation.
In the hadronic analysis, an additional uncertainty is used to account for small differences in b tagging which arise from the larger jet-cone size.
Details on the systematic uncertainty in b tagging can be found in Section~\ref{sec:syst}.

\subsubsection{Tagging of top quarks\label{sec:ttag}}

The large Lorentz boost of the top quark from heavy $\PWRpr$ boson ($M_{\PWRpr}\gtrsim1\TeV$) decays causes the three jets from hadronic decays to merge into a single large-radius jet with distinct substructure.
Variables that are sensitive to characteristics of this substructure can be used to discriminate signal from background.
The hadronic analysis uses a top tagging algorithm that is based on three such variables: jet mass, N-subjettiness \cite{Nsubjet1,Nsubjet2}, and subjet b tagging.

The jet mass is calculated after applying the modified mass-drop tagger, also known as the ``soft drop" algorithm \cite{softdrop,Larkoski2014}, which
reclusters the AK8 jet with the Cambridg--Aachen algorithm~\cite{Dokshitzer:1997in} and declusters until the following requirement is met:
\begin{equation}
 \frac{\min(p_{\mathrm{T_1}},p_{\mathrm{T_2}})}{p_{\mathrm{T_1}}+p_{\mathrm{T_2}}}>z(\Delta R_{\mathrm{12}}/R_\mathrm{0})^\beta ,
\end{equation}
where $p_{\mathrm{T_i}}$ are the magnitude of the transverse momenta of the two subjet candidates, $\Delta R_{12}$ is the distance ($\Delta R = \sqrt{\smash[b]{(\Delta\eta)^2+(\Delta\phi)^2}}$, where $\phi$ is the azimuthal angle in radians) between candidates, and $R_0$ is the jet size parameter.
For this analysis, we use $z = 0.1$ and $\beta = 0$, and require the mass of the soft-drop declustered jet to be between 110 and 210\GeV, i.e. consistent with the top quark mass, $M_\mathrm{top}$.
For this operating point, the soft drop algorithm is equivalent to the modified mass-drop tagger~\cite{Dasgupta:2013via,softdrop}.

The N-subjettiness algorithm defines a series of $\tau_\mathrm{N}$ variables that describe the consistency between the jet energy and the number of assumed subjets (N):
\begin{equation}
 \tau_\mathrm{N}=\frac{1}{d}\sum_i p_{\mathrm{T_i}}\min(\Delta R_{\mathrm{1,i}},\Delta R_{\mathrm{2,i}},\ldots,\Delta R_{\mathrm{N,i}}) ,
\end{equation}
where $\Delta R_{J,i}$ is the distance between the axis of the subjet candidate ($J$) and a specific constituent particle ($i$), and $d$ is the normalization factor,
\begin{equation}
 d=\sum_i p_{\mathrm{T_i}}R ,
\end{equation}
where $R$ is the distance parameter used in the jet clustering algorithm.
The axes of the subjet candidate used to calculate N-subjettiness are found using the exclusive \kt algorithm~\cite{Nsubjet3}, after which an optimization procedure is
applied to minimize the N-subjettiness value, calculated using all particle-flow constituents of the AK8 jet.
A jet with a low $\tau_\mathrm{N}$ value will have energy deposited close to the axes of the N subjet candidates, which is a characteristic of a jet containing N subjets.
A top quark jet is likely to be more consistent with three subjets than two, while a jet from a gluon or light quark will typically be consistent with either two or three subjets.
Therefore, the ratio of $\tau_\mathrm{3}$ and $\tau_\mathrm{2}$ is characteristically smaller for top quark jets than for the multijet background.
We select jets with $\tau_\mathrm{3}/\tau_\mathrm{2} < 0.61$.

Finally, we apply the CSVv2 b tagging algorithm to the two soft-drop subjets of the candidate jet, and
require the maximum b tagging discriminator value $(SJ_\textrm{b tag})$ to be at least 0.76.  The above selection criteria correspond to the working point of the CMS top quark tagging algorithm defined by a 0.3\% top-quark mistagging rate~\cite{JME-15-002}, with a corresponding top-quark efficiency of approximately 30\%.

Scale factors resulting from small differences in t tagging efficiencies in data and simulation are derived in a pure semileptonic \ttbar sample separately for jets with \pt greater or less than 550\GeV.
These are applied as corrections to simulated events, and are consistent with unity.

\subsection{Identification of electrons and muons \label{sec:lepid}}

Electron candidates are selected using a multivariate identification technique, specifically, a boosted decision tree. The multivariate discriminant is based on the spatial energy distribution of the shower, the quality of the track, the match between the track and electromagnetic cluster, the fraction of total cluster energy deposited in the HCAL, the amount of energy appearing in the regions surrounding the tracker and calorimeters, and the probability of the electron to have originated from a converted photon.
The track associated with a muon candidate is required to have hits in the pixel and muon detectors, good quality, and transverse and longitudinal impact parameters (distance of closest approach) with respect to the leading vertex close to zero.

Both the leptonic and hadronic analyses use the same criteria for muon identification, while the criteria used for electron identification are less restrictive in the hadronic analysis than in the leptonic analysis.
The choice of lepton identification and use of a veto ensure that there is no overlap between events in the two analyses, and makes combining their results straightforward.

Scale factors arising from small differences between lepton identification efficiencies in data and simulation are obtained from a data sample of $\Z\to\ell\ell$ events as a function of $\abs{\eta}$.
These scale factors are then applied as corrections to simulated events.

In highly boosted semileptonic top quark decays from heavy $\PWRpr$ bosons, the lepton and jet may not be well separated.
For this reason, no isolation requirement is applied to the lepton.
Instead, a two-dimensional requirement is placed on the $\Delta R$ and $\pt^\textrm{rel}$ for the lepton and the closest jet with $\pt >25$\GeV and $\abs{\eta}<2.5$, where the $\pt^\text{rel}$ is given by the magnitude of the component of the lepton momentum orthogonal to the jet axis.
For electrons (muons), we require that either $\Delta R>0.4$ or $\pt^\text{rel}>60 (50)$\GeV.
These requirements help remove the multijet contribution from the background in the leptonic analysis, while maintaining high efficiency for signal events.
The four-momenta of identified lepton-candidate particles are subtracted from the four-momentum of the jets that contain them,
which helps ensure that jets considered in the leptonic analysis are not contaminated by nearby high-energy leptons.

\subsection{Mass reconstruction}

The methods of reconstructing $\PWRpr$ boson candidates differ in the two analyses.  In the leptonic channel, the tb invariant mass
is reconstructed from the charged
lepton, \ptvecmiss, and two jets in the event. The $x$- and $y$-components of neutrino \pt are determined from \ptvecmiss and the $z$-component is calculated by constraining the invariant mass of the
lepton and neutrino to the mass of the W boson. This leads to a quadratic equation in $p^\nu_z$.
When the two solutions are real numbers, both are used to reconstruct W boson
candidates. If both solutions contain imaginary parts, we set $p^\nu_z$ to the real part of the
solutions, and recompute $p^\nu_\mathrm{T}$, which yields a different quadratic ambiguity. In the latter case, we
use only the solution with mass closest to 80.4\GeV.
Once we have all components of the neutrino momentum, we combine the viable neutrino momentum solutions with the charged lepton momentum to create W boson candidates. We then reconstruct
the top quark by combining the four-momenta of each of the W boson candidates with each jet with $\pt>25$\GeV and $\abs{\eta}<2.4$.
Whichever jet yields a top quark candidate mass closest to 172.5\GeV is
labeled as the ``best jet`` and is used to reconstruct the top quark
candidate. In the case of two W candidates, we use the candidate that yields the top quark
mass closest to its nominal value of 172.5\GeV. Finally, we combine the top quark candidate with the highest \pt jet, that is not the ``best jet,''
yielding the reconstructed $\PWRpr$ candidate.

In the hadronic channel, the tb invariant mass is reconstructed from the two leading AK8 jets in the event.

\subsection{Analysis selections in the leptonic channel}

Candidate events in the leptonic analysis are selected in the HLT with single-lepton triggers that require a \pt of at least 105 (45)\GeV for electrons (muons) and have no isolation requirement.
Scale factors to account for differences in efficiency between data and simulation are obtained through the procedure outlined in Section~\ref{sec:lepid}.
Events must contain a reconstructed lepton with $\pt>180$\GeV and $\abs{\eta}< 2.5 (2.1)$ in the electron (muon) channel.
Events are rejected if they contain more than one identified lepton with $\pt>35$\GeV and $\abs{\eta}< 2.5 (2.1)$ in the electron (muon) channel.

Events are also required to have at least two jets with $\pt>30$\GeV and $\abs{\eta}<2.4$, and the jet with leading \pt must have $\pt>350 (450)$\GeV in the electron (muon) channel, where at
least one of these jets must be b tagged.
Events must have $\ptvecmiss>120 (50)$\GeV in the electron (muon) channel.
In addition, events in the electron channel must have an opening angle in the transverse plane between the electron and the \ptvecmiss vector $|\Delta\phi(\mathrm{e},\ptvecmiss)|<2$ radians.
In both channels, the top quark candidate is required to have $\pt^{\PQt}>$ 250\GeV and $\pt^{j_1+j_2}>$ 350\GeV, where $\pt^{j_1+j_2}$ is the \pt of the vector sum of the two leading \pt jets.
In addition, in the muon channel, the mass of the top quark candidate must satisfy the condition 100 $<m_{\PQt}<$ 250\GeV.
These requirements all serve to reject events which are not consistent with the decay of a heavy resonance to a top and bottom quark.
The selections in both channels are optimized separately, thereby leading to slight differences in certain requirements.
Event yields after the selection for the leptonic analysis are shown in Table~\ref{table:lep_yields}.

\begin{table}[htb]
\centering
\topcaption{Number of selected events, and the number of signal and background events expected from simulation in the leptonic analysis.
The expectations for signal and background correspond to an integrated luminosity of 2.2\fbinv.
``Full selection'' refers to the additional requirements of $\pt^{\PQt}>250$\GeV and $\pt^{j_1+j_2}>350$\GeV for both channels, and also 100 $<m_{\PQt}<$ 250\GeV in the muon channel, while ''Object selection'' omits these requirements.
The quoted uncertainty does not include systematic uncertainties that affect the shape of distributions (a complete description of sources of uncertainty can be found in Section~\ref{sec:syst}).}
\label{table:lep_yields}
\resizebox{\textwidth}{!}{
\begin{tabular}{lcccccccc}
\hline
 & \mc{4}{c|}{Electron channel} & \mc{4}{c}{Muon channel} \\
 & \mc{2}{c|}{Object selection} & \mc{2}{c|}{Full selection} & \mc{2}{c|}{Object selection} & \mc{2}{c}{Full selection}\\
 & \mc{1}{c|}{$1$ b tag} & \mc{1}{c|}{$2$ b tags} & \mc{1}{c|}{$1$ b tag} & \mc{1}{c|}{$2$ b tags} & \mc{1}{c|}{$1$ b tag} & \mc{1}{c|}{$2$ b tags} & \mc{1}{c}{$1$ b tag} & \mc{1}{c}{$2$ b tags}\\\hline
{Signal} & \mc{8}{c}{} \\\hline
$M_{\PWRpr}$ = 1400\GeV & \mc{1}{c}{30} & \mc{1}{c}{22} & \mc{1}{c}{28} & \mc{1}{c}{20} & \mc{1}{c}{35} & \mc{1}{c}{31} & \mc{1}{c}{26} & \mc{1}{c}{24}\\
$M_{\PWRpr}$ = 2000\GeV & \mc{1}{c}{9} & \mc{1}{c}{6} & \mc{1}{c}{9} & \mc{1}{c}{6} & \mc{1}{c}{11} & \mc{1}{c}{9} & \mc{1}{c}{9} & \mc{1}{c}{7}\\
$M_{\PWRpr}$ = 2600\GeV & \mc{1}{c}{3} & \mc{1}{c}{1} & \mc{1}{c}{3} & \mc{1}{c}{1} & \mc{1}{c}{3} & \mc{1}{c}{2} & \mc{1}{c}{3} & \mc{1}{c}{1}\\
\hline\\[-2ex]
{Background} & \mc{8}{c}{} \\\hline
\ttbar & \mc{1}{c}{71} & \mc{1}{c}{26} & \mc{1}{c}{56} & \mc{1}{c}{19} & \mc{1}{c}{68} & \mc{1}{c}{27} & \mc{1}{c}{49} & \mc{1}{c}{18}\\
tqb & \mc{1}{c}{5} & \mc{1}{c}{2} & \mc{1}{c}{4} & \mc{1}{c}{1} & \mc{1}{c}{4} & \mc{1}{c}{1} & \mc{1}{c}{3} & \mc{1}{c}{1}\\
tW & \mc{1}{c}{11} & \mc{1}{c}{6} & \mc{1}{c}{10} & \mc{1}{c}{5} & \mc{1}{c}{9} & \mc{1}{c}{3} & \mc{1}{c}{4} & \mc{1}{c}{1}\\
$\PAQt$W & \mc{1}{c}{11} & \mc{1}{c}{4} & \mc{1}{c}{9} & \mc{1}{c}{4} & \mc{1}{c}{9} & \mc{1}{c}{4} & \mc{1}{c}{5} & \mc{1}{c}{2}\\
tb & \mc{1}{c}{1} & \mc{1}{c}{0} & \mc{1}{c}{1} & \mc{1}{c}{0} & \mc{1}{c}{0} & \mc{1}{c}{0} & \mc{1}{c}{0} & \mc{1}{c}{0}\\
$\PW(\to\ell\nu)$+jj & \mc{1}{c}{89} & \mc{1}{c}{8} & \mc{1}{c}{77} & \mc{1}{c}{7} & \mc{1}{c}{80} & \mc{1}{c}{6} & \mc{1}{c}{25} & \mc{1}{c}{1}\\
$\PW(\to\ell\nu)$+bb/cc & \mc{1}{c}{139} & \mc{1}{c}{22} & \mc{1}{c}{119} & \mc{1}{c}{18} & \mc{1}{c}{128} & \mc{1}{c}{23} & \mc{1}{c}{45} & \mc{1}{c}{7}\\
$(\Z\to\ell\ell)$+jets & \mc{1}{c}{3} & \mc{1}{c}{0} & \mc{1}{c}{4} & \mc{1}{c}{0} & \mc{1}{c}{21} & \mc{1}{c}{0} & \mc{1}{c}{12} & \mc{1}{c}{0}\\
WW, WZ, ZZ & \mc{1}{c}{9} & \mc{1}{c}{0} & \mc{1}{c}{7} & \mc{1}{c}{0} & \mc{1}{c}{3} & \mc{1}{c}{0} & \mc{1}{c}{0} & \mc{1}{c}{0}\\
\hline{Total background} & \mc{1}{c}{339$\pm$22} & \mc{1}{c}{67$\pm$5} & \mc{1}{c}{287$\pm$19} & \mc{1}{c}{53$\pm$4} & \mc{1}{c}{322$\pm$24} & \mc{1}{c}{64$\pm$5} & \mc{1}{c}{143$\pm$11} & \mc{1}{c}{30$\pm$3}\\
\hline\\[-2ex]
{Data} & \mc{1}{c}{309} & \mc{1}{c}{58} & \mc{1}{c}{256} & \mc{1}{c}{44} & \mc{1}{c}{281} & \mc{1}{c}{58} & \mc{1}{c}{143} & \mc{1}{c}{30}\\
\hline
\end{tabular}}
\end{table}

\subsection{Analysis selections in the hadronic channel}

Candidate events in the hadronic channel are required to satisfy one of two HLT selections.
The first demands at least two AK8 jets with $\pt>200$\GeV, one of which must have a trimmed \cite{Krohn2010} jet mass greater than 30\GeV, and also requires the leading \pt jet to have $\pt>280$\GeV.
In addition, this trigger requires that the event contains at least one b-tagged jet.
The second trigger requires that the scalar \pt sum of reconstructed jets be at least 800\GeV.
The efficiency of the combination of these two triggers is measured with data collected using a trigger with a lower scalar \pt sum threshold, and
is extracted as a function of the scalar \pt sum of the two jets with leading \pt ($H_\mathrm{T}$), which provides a way to account for this effect.

We require events to have at least two jets with $\pt>350$\GeV,
one of which must be identified as a top jet using the t tagging algorithm, and the other must be tagged as a bottom jet.
Furthermore, the b jet must have a soft-drop mass less than 70\GeV.
Finally, the two jets are required to be separated by $\abs{\Delta\phi}>\pi/2$ radians and to have $\abs{\Delta y}<1.3$, where $\Delta y$ is the rapidity difference between the two jets.

The event yields after implementing the selections in the hadronic analysis are shown in Table~\ref{table:had_yields}.

\begin{table}[htb]
\centering
\topcaption{Number of selected events, and the number of signal and background events expected from simulation in the hadronic analysis.
The expectations for signal and background correspond to an integrated luminosity of 2.6\fbinv.
The quoted uncertainty does not include systematic uncertainties that affect the shape of distributions (a complete description of sources of uncertainty can be found in Section~\ref{sec:syst}).}
\label{table:had_yields}
\begin{tabular}{lc}
\hline
{Signal} & \\ \hline
$M_{\PWRpr}$ = 1400\GeV & 228 \\
$M_{\PWRpr}$ = 2000\GeV & 27  \\
$M_{\PWRpr}$ = 2600\GeV & 4  \\
\hline\\[-2ex]
{Background} & \\ \hline
Multijets & 6134\\
\ttbar & 376 \\
tW & 32\\
\hline{Total background} & $6542 \pm102$ \\
\hline\\[-2ex]{Data} & 6491 \\
\hline
\end{tabular}
\end{table}

\section{Backgrounds\label{sec:anstrat}}

\subsection{Backgrounds in the leptonic analysis}

\subsubsection{Top quark pair production background}

The predicted \ttbar background is estimated from simulation and checked in two distinct control regions, both of which do not apply the requirements on $p_\mathrm{T}^{j_1+j_2}$, $p_\mathrm{T}^{\PQt}$, $m_{\PQt}$, nor the number of b jets.
The first region is defined by relaxing the leading jet \pt and \ptvecmiss requirements, and requiring events to have at least four jets, two of which are b-tagged, and have $400<\mtb<750$\GeV.
The latter requirement ensures that the signal contamination in this region is less than 1\%.
The second region is defined by requiring events to have two leptons, which must have $\pt>150 (35)\GeV$ for the leading (subleading) \pt lepton.
This requirement ensures that there is no overlap between the signal region and the second control region.
In addition, we relax the requirements on the leading jet \pt and \ptvecmiss, and reject events for which the invariant mass of the dilepton system (if they are of the same flavor) is between 70 and 110\GeV,
which ensures that the control region does not contain a significant fraction of $\Z/\gamma^*$+jets events.

In both control regions, we compare simulated distributions and overall yields with data.
We observe significantly better agreement between data and simulation when a correction is applied to the top quark \pt spectrum in the \ttbar simulation.
The correction factor is obtained from measurements of the differential top quark \pt distribution~\cite{Khachatryan:2015oqa}.
We apply this correction factor to the \ttbar simulation, as a function of the generator-level top quark \pt, and use the differences from the distributions without the correction as estimates of the systematic uncertainty in the expected \ttbar background.

\subsubsection{W+jets background}

The prediction for the W+jets background is estimated from simulation.
It is then corrected for known discrepancies in the relative fraction of W+jets events with light-flavor jets compared to bottom or charm quark jets.
This correction is obtained from data using a modified event selection that does not include the requirements on $p_\mathrm{T}^{j_1+j_2}$, $p_\mathrm{T}^{\PQt}$, and $m_{\PQt}$, and also removes the requirement of a b-tagged jet.
This sample is referred to as the pre-tag sample.
A subset of these events, in which neither of the two leading \pt jets are b tagged, is referred to as the 0-tag sample.
The 0-tag sample is dominated by the W+jets background and contains contributions from other background sources, which comprise less than 20\% of the total.
The difference between data and simulation in the 0-tag sample is used to obtain a first-order scale factor for W+jets light-flavor events, which is applied to the W+jets
simulation, and the difference between data and simulation in the pre-tag distribution is used to calculate a first-order scale factor for W+jets heavy-flavor events.
This procedure is repeated until following iterations do not cause the scale factors to shift by more than 0.1\%.
We also check this calculation
by analytically solving the system of equations from the iteration, and confirm that the two methods yield identical results.

We require that the total number of predicted events is unaffected by the simultaneous application of the two scale factors.  We assign uncertainties to these factors by repeating the procedure with the b tagging scale factors varied within their uncertainties.
The procedure is identical to the procedure used in Ref.~\cite{Chatrchyan:2014koa}.

\subsection{Backgrounds in the hadronic analysis}

\subsubsection{Multijet background}

The multijet background is estimated from data,
and the method is verified through simulation.
The procedure uses the distribution of multijet events that fail the b tagging requirement,
weighted by a transfer factor (average b tagging rate) to predict the multijet yield in the signal region.

To estimate the average b tagging rate in multijet events, we define modified t tagging criteria.
Specifically, we now select events that contains jets with $\tau_\mathrm{3}/\tau_\mathrm{2}~>~$0.75, and shift the soft-drop jet mass window to be between 50 and 170\GeV.
These requirements ensure that the control region is orthogonal to the signal region and has contributions from both signal and \ttbar events that are less than 1\%.
Using the standard $SJ_\textrm{b tag}$ requirement in the signal region, we favor a similar parton flavor composition.
A control region is then defined by applying the signal selection with the modified t tagging requirements, omitting the b tagging requirement.

We calculate the average b tagging rate as a function of b candidate jet \pt in three $\abs{\eta}$ regions: $\abs{\eta}<0.50$ (low), $0.50\leq\abs{\eta}<1.15$ (transition), $1.15\leq\abs{\eta}<2.40$ (high).
The denominator contains all events in the control region, while the numerator includes only those that pass the signal region b tagging requirement.
The average b tagging rate in each $\abs{\eta}$ range is fitted using a bifurcated polynomial that models the distribution.
The functional form is
\begin{equation}
 f(\pt)=\left\{
\begin{aligned}
c_0+c_1\pt+c_2(\pt-a)^2,&\quad\text{if }\pt<a\\
c_0+c_1\pt+c_3(\pt-a)^2,&\quad\text{if }\pt\ge a.
 \end{aligned}
 \right.
\end{equation}
The parameters $c_0$ to $c_3$ are free coefficients determined in the fit.
The value of $a$ is chosen separately for each $\abs{\eta}$ region, and is 500, 500, and 550\GeV in the low, transition, and high-$\abs{\eta}$ regions, respectively.

The uncertainty related to the average b tagging rate is obtained from the full covariance matrix of the fitting algorithm.
The functional form is chosen to optimize agreement between sideband and Monte Carlo estimates.
We estimate an uncertainty related to the choice of the fit function by comparing the results of
the nominal fit with those determined using other functional forms.
These other forms
include the following: a constant, a second-degree polynomial, a
third-degree polynomial, and an exponential function.

We observe that there is a correlation between the b tag rate and the soft-drop mass of the b candidate.
To account for this correlation, we extract a correction factor for the multijet background as a function of the soft-drop mass of the b jet candidate.
This factor is calculated by taking the ratio of the soft-drop mass distributions for the b tagging pass and b tagging fail samples in the control region of the multijet simulation.
The factor is then used as an event weight along with the fit to the average b tagging rate to estimate the multijet background from data.
An uncertainty in the factor, equal to half the difference between the factor and unity, is included in the analysis.

We check the closure of this procedure using both multijet simulation and an additional control region in data.
The control region is defined by inverting the $SJ_\textrm{b tag}$ requirement in the signal region.
This provides a much purer multijet sample in data, which is orthogonal to both the signal region and the control region used to estimate the multijet contribution.

The closure test using the prediction from simulation shows a small residual discrepancy in the \mtb distribution, which is used to extract a correction for the multijet prediction.
We include an uncertainty in this correction equal to the difference between the correction and unity.
After this correction, the corresponding closure test in the data control region shows good agreement between the multijet prediction and
observed data.

\subsubsection{Top quark pair production background}

In the hadronic analysis, the \ttbar background prediction is estimated from simulation and checked in a region defined through selections identical to those used in the signal region, except that the b jet soft-drop mass requirement is inverted.
This region contains an increased fraction of \ttbar events relative to the signal region (approximately a factor of six), and does not overlap with the signal region or any other control regions used in the analysis.
The prediction for the multijet background in this region is estimated from data using the same method as the signal region.
The prediction for the \ttbar background is found to be consistent with that observed in the data, and no other correction is required.
\section{Systematic uncertainties\label{sec:syst}}

Systematic uncertainties fall into two categories: those that affect only the total event yield, and those that affect both the event yield and the \mtb distribution.
Unless otherwise specified, the uncertainties are common both the leptonic and hadronic analyses.

The uncertainty in the measured integrated luminosity (2.7\%)~\cite{2015lumi} belongs to the first category.
The leptonic analysis includes uncertainties on the modeling  of the lepton trigger (2-4\%).
The hadronic analysis includes uncertainties in the AK4 vs. AK8 jet b tagging rates (3\%), t tagging efficiency (20\%)
, and in the theoretical \ttbar and single top quark cross sections ($\approx5\%$).

Since the two analyses use the same criteria to identify muons, but different criteria for electrons, the uncertainty in the muon reconstruction and identification (2\%) is included in both analyses, while the uncertainty in electron reconstruction and identification (5\%) is included only in the leptonic analysis.

Other uncertainties belong to the second category and are detailed below.
Unless otherwise specified, the uncertainties are assigned to all samples for which the prediction is estimated from simulation.

The uncertainties due to the choice in the renormalization and factorization scales ($\mu_R$ and $\mu_F$, respectively) are evaluated at the matrix element level using event weights to change the scales up or down relative to the nominal scale by a factor of two, while restricting to $0.5\le\mu_R/\mu_F\le2$~\cite{Cacciari:2003fi,Catani:2003zt}.
The uncertainty from changes in both scales at the parton shower level are evaluated for the \ttbar background using samples generated with twice or half the nominal scale.

Uncertainties on the b tagging, jet energy scale, and jet energy resolution are calculated by varying the relevant scale factors within their uncertainties.
For the jet energy scale and resolution, nominal factors and uncertainties are obtained for both AK4 and AK8 jets and applied appropriately in the leptonic and hadronic analyses.

A correction is applied to all simulated event samples to provide better matching of the distribution of pileup interactions in data.
This procedure uses a minimum bias interaction cross section ($\sigma_\mathrm{mb}$) of 69\unit{mb}, and uncertainties are calculated by varying the minimum bias cross section by $\pm5\%$.

To estimate the uncertainty arising from the choice of the PDF,
we use the NNPDF 3.0 PDF set uncertainty defined in Ref.~\cite{Botje:2011sn}.

In the leptonic analysis, the uncertainties in the W+jets heavy- and light-flavor factors are included as a variation in the W+jets background, and the \ttbar background with an uncorrected top quark \pt spectrum is included as a one-sided systematic uncertainty.

In the hadronic analysis, the uncertainty in the trigger efficiency is taken to be one half of the measured trigger inefficiency, and applied as a function of the scalar \pt sum of the two leading jets.
Uncertainties in the multijet background estimation procedure are also applied.
These result from choice of functional form in the fit to the average b tagging rate, corrections due to correlations between the average b tagging rate and soft-drop jet mass,
and differences
obtained from a closure test in
simulation.

In the leptonic analysis, the dominant uncertainty sources are from the correction to the \pt spectrum of the top quark in \ttbar events, and $\mu_R$ and $\mu_F$ at the matrix element level.
In the hadronic analysis, the dominant uncertainty sources are from the multijet background estimation and t tagging efficiency.
Both analyses are also affected by the subdominant uncertainties related to the choice of PDF and b tagging.
All systematic uncertainties for both analyses are summarized separately in Table~\ref{table:syst}.

\begin{table}[htb]
\topcaption{Sources of systematic uncertainty affecting the $\mtb$ distribution taken into account when setting 95\% CL upper limits.
The three right-most columns indicate the channels to which the uncertainty applies (noted by $\circ$), and whether it also applies to signals (noted by $\checkmark$).
When a source applies to both channels, it is treated as fully correlated in the combination.
Sources that list the changes as $\pm$1 standard deviation (s.d.) depend on the distribution of the variable given in the parentheses, while those that list the variation as a percent are rate uncertainties.}
\centering
\begin{tabular}{llccc}
\hline
{Source} & {Variation}  & {Leptonic}  & {Hadronic} & {Signal}\\
\hline
Integrated luminosity & $\pm$2.7\% & $\circ$ & $\circ$ & \checkmark \\
Muon identification efficiency  & $\pm$2\% & $\circ$ & $\circ$ & \checkmark\\
Electron identification efficiency  & $\pm$5\% & $\circ$ & & \checkmark\\
Single-lepton trigger ($\Pe/\mu$) & $\pm$4\%/2\% & $\circ$ & & \checkmark\\
AK4 to AK8 b tagging & ${\pm}$3\%  & & $\circ$ &\checkmark\\
Top quark tagging & $\pm$20\%  & & $\circ$ & \checkmark \\
$\ttbar$ cross section & ${+}4.8$\%, ${-}5.5$\% & & $\circ$ & \\
tW cross section & $\pm$5.4\% & & $\circ$ & \\
Matrix element $\mu_R/\mu_F$ scales & ${\pm}1\mathrm{s.d.} (\mu_R/\mu_F)$ & $\circ$ & & \\
\ttbar parton shower scale & ${\pm}1\mathrm{s.d.} (\mu_R/\mu_F)$ & $\circ$ & $\circ$ & \\
Jet energy scale & ${\pm}1\mathrm{s.d.} (\pt,\eta)$ & $\circ$ & $\circ$ & \checkmark\\
Jet energy resolution & ${\pm}1\mathrm{s.d.} (\pt, \eta)$ & $\circ$ &  $\circ$ & \checkmark\\
b tagging & ${\pm}1\mathrm{s.d.} (\pt)$ & $\circ$ & $\circ$ & \checkmark \\
Light quark mistag rate & ${\pm}1\mathrm{s.d.}(\pt,\eta)$ & $\circ$ & & \checkmark\\
Pileup & ${\pm}1\mathrm{s.d.}$ ($\sigma_\mathrm{mb}$) & $\circ$ &  $\circ$ & \checkmark\\
PDFs & ${\pm}1\mathrm{s.d.}$ &  $\circ$ & $\circ$ & \checkmark \\
W+jets heavy-flavor fraction & ${\pm}1\mathrm{s.d.}$ & $\circ$ & & \\
Top $\pt$ reweighting & $+1\mathrm{s.d.}$ & $\circ$ & & \\
$H_\mathrm{T}$ trigger & ${\pm}1\mathrm{s.d.} (H_\mathrm{T})$ & & $\circ$ & \checkmark\\
Average b tagging rate fit & ${\pm}1\mathrm{s.d.} (\pt, \eta)$ & & $\circ$ & \\
Alternative functional forms & ${\pm}1\mathrm{s.d.} (\pt, \eta)$ & & $\circ$ & \\
b candidate mass & ${\pm}1\mathrm{s.d.} (M_{\PQb})$ & & $\circ$ & \\
Multijet simulation nonclosure & ${\pm}1\mathrm{s.d.} (\mtb)$ & & $\circ$ & \\
\hline
\end{tabular}

\label{table:syst}
\end{table}
\section{Results\label{sec:results}}

\begin{figure}[htb]
  \centering
    \includegraphics[width=0.48\textwidth]{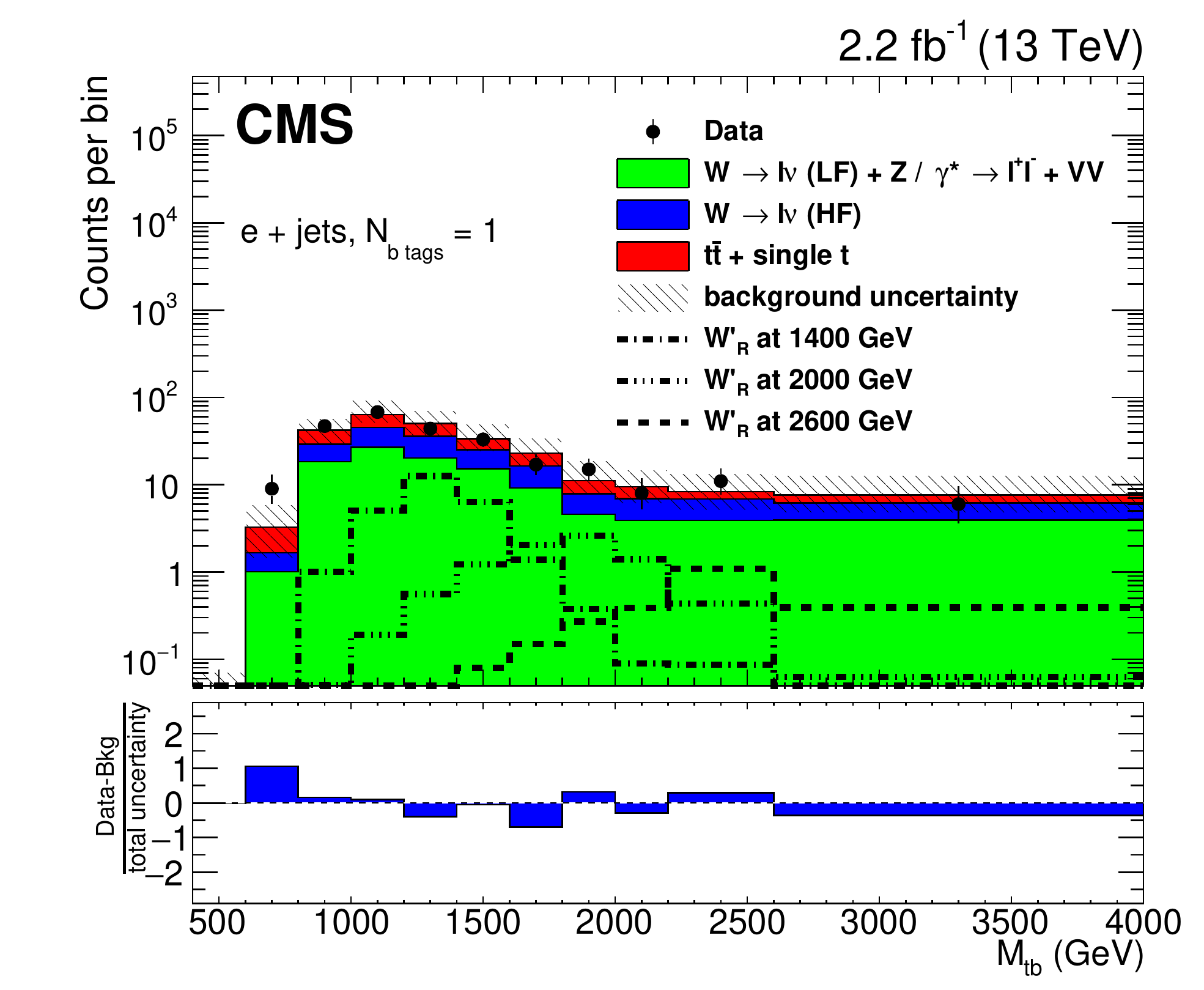}
    \includegraphics[width=0.48\textwidth]{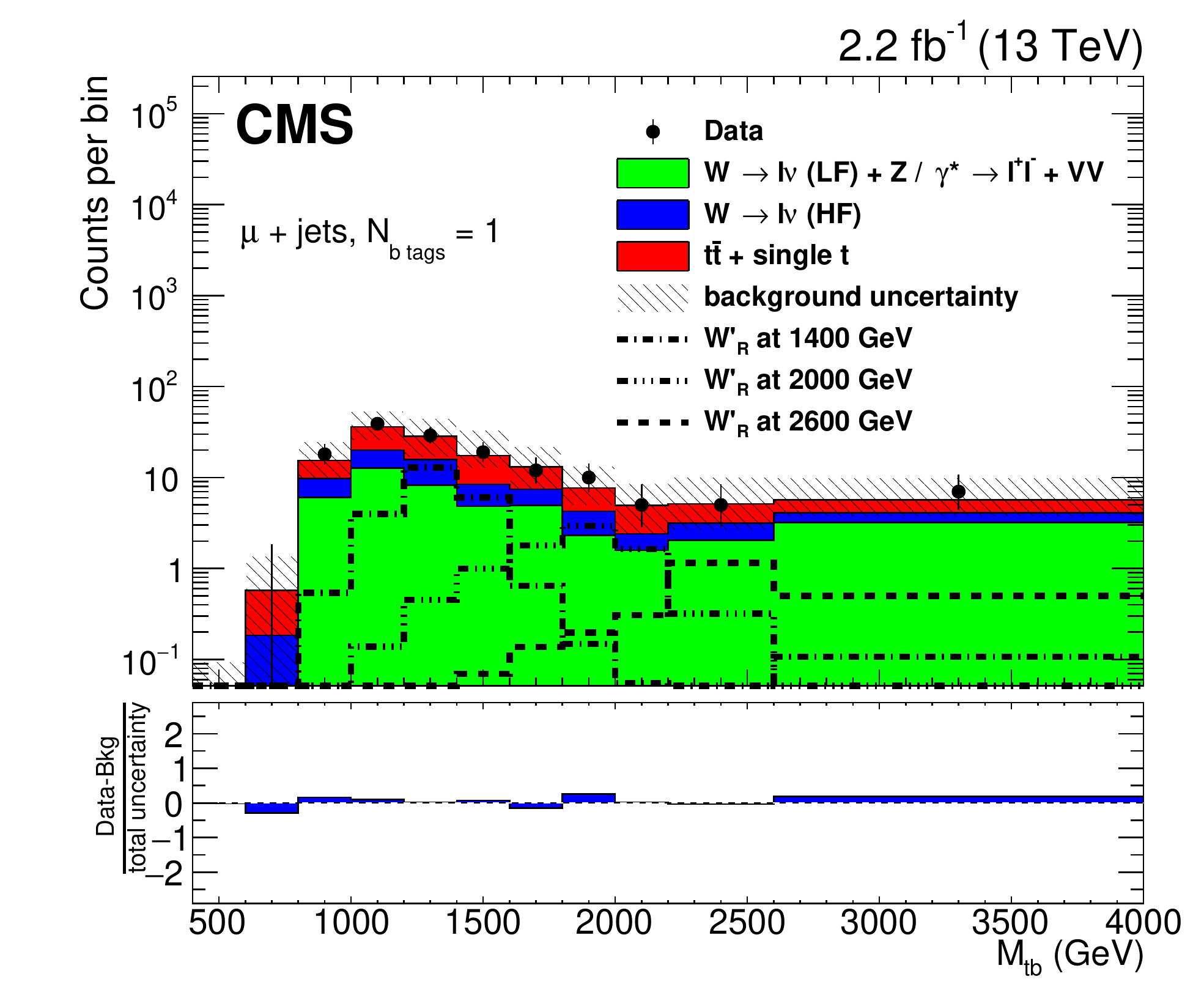}
    \includegraphics[width=0.48\textwidth]{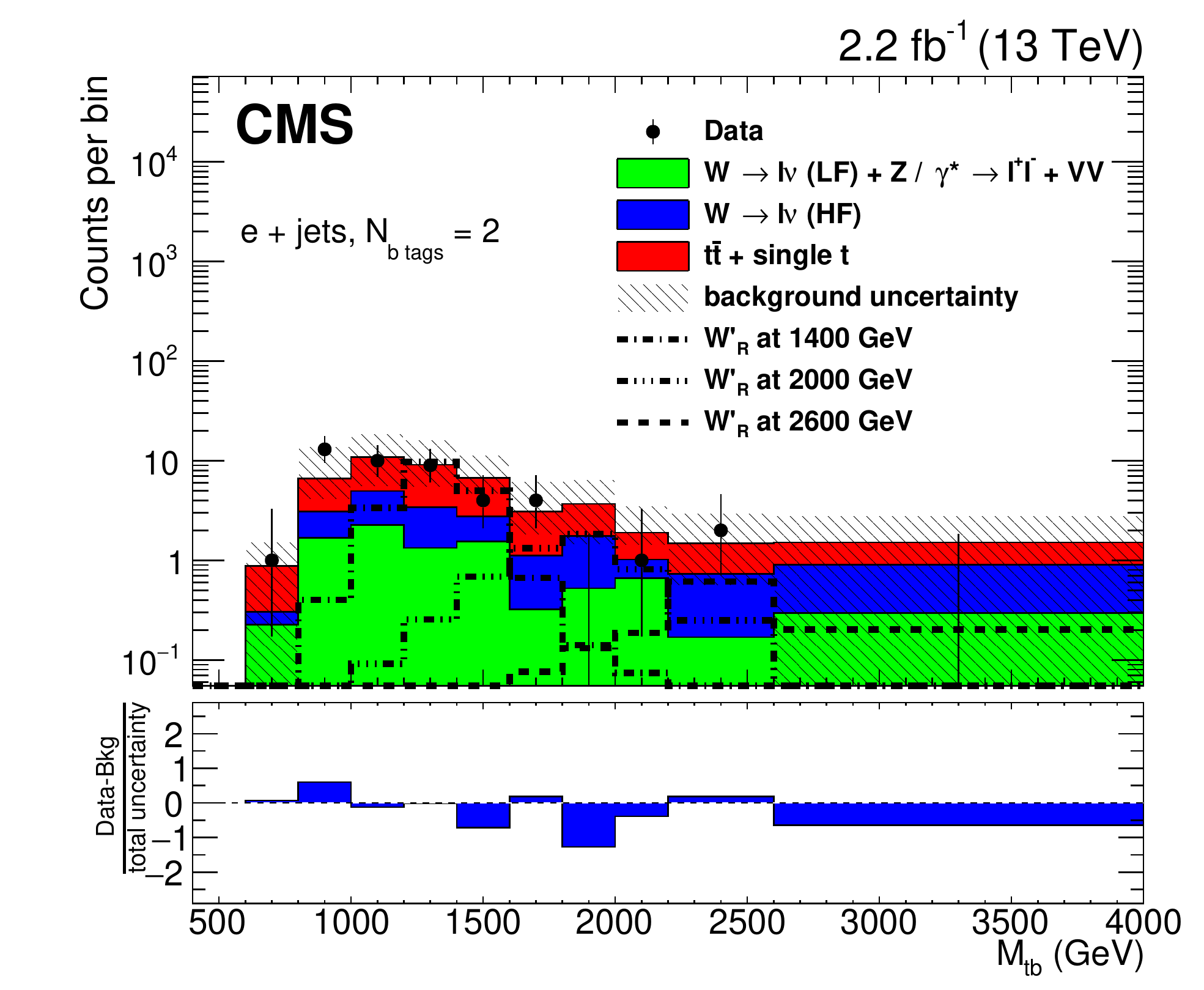}
    \includegraphics[width=0.48\textwidth]{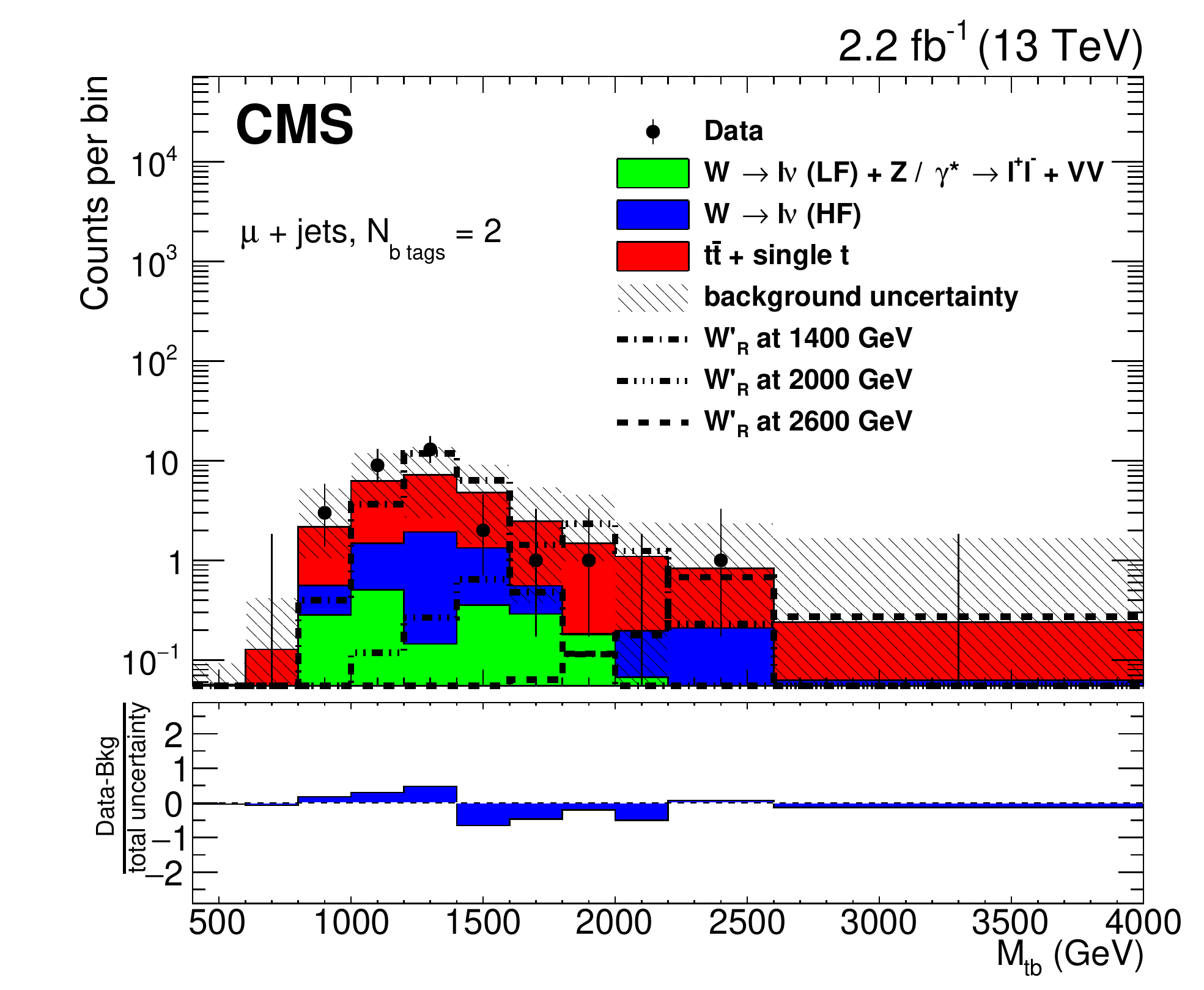}
    \caption{Reconstructed \mtb distributions from the leptonic analysis in the 1 b tag (upper) and 2 b tag (lower) categories, for the electron (left) and muon (right) channels.
The ``LF'' and ``HF'' notations indicate the light- and heavy-flavor components of the W+jets contribution, respectively.
The simulated $\PWRpr$ signal and background samples are normalized to the integrated luminosity of the analyzed data set.
The distributions are shown after the application of all selections.
The 68\% uncertainty in the background estimate includes all contributions to the predicted background, while the total uncertainty is the combined uncertainty of the background and data.
}
    \label{fig:lep_mtb}

\end{figure}

\begin{figure}[htb]
\centering
\includegraphics[width=0.7\textwidth]{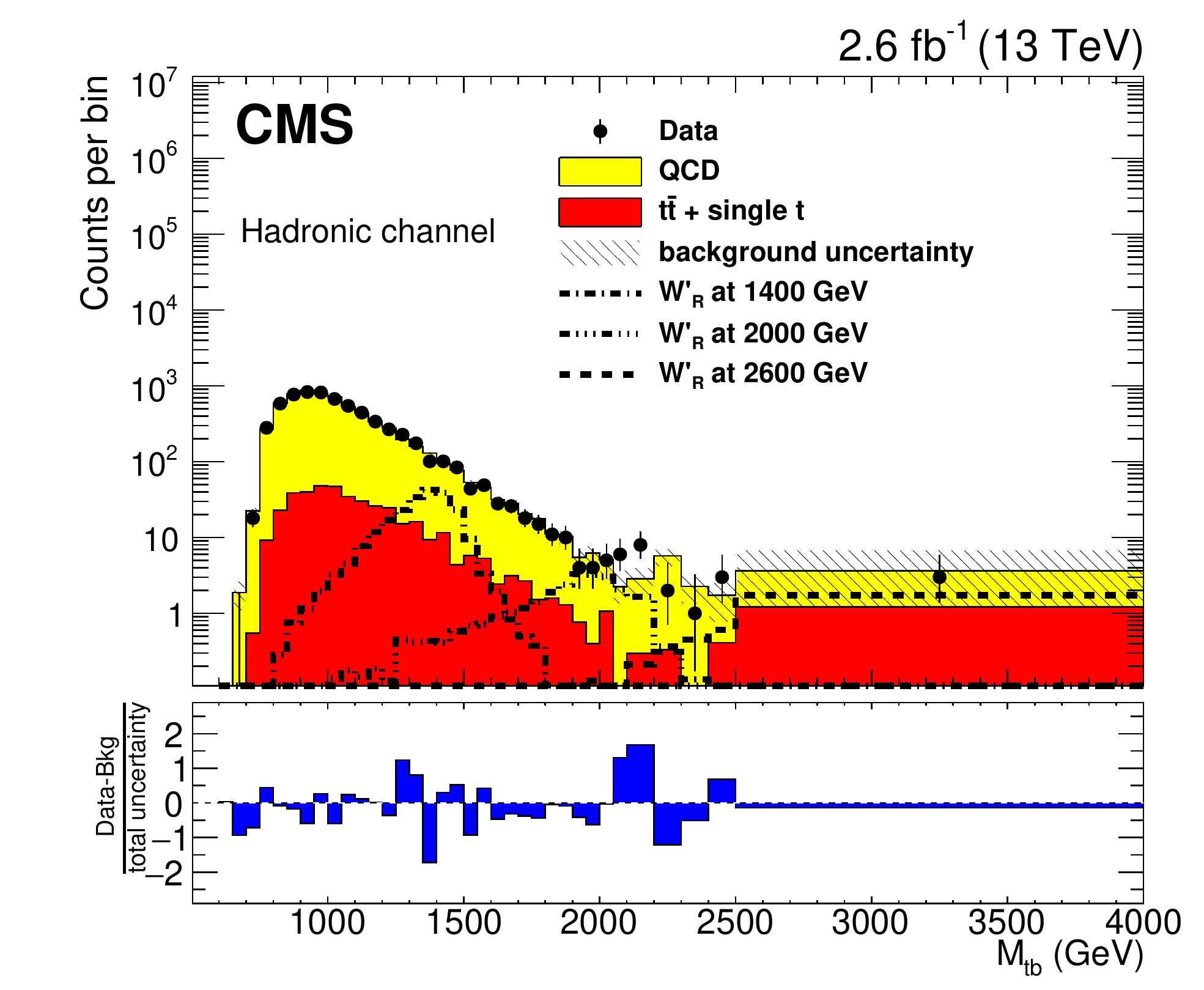}
\caption{
Reconstructed \mtb distribution from the hadronic analysis.
The simulated $\PWRpr$ signal and backgrounds are normalized to the integrated luminosity of the analyzed data set.
The distribution is shown after the application of all selections.
The 68\% uncertainty in the background estimate includes all contributions to the predicted background, while the total uncertainty  is the combined uncertainty of the background and data.
}
\label{fig:had_mtb}

\end{figure}

Comparisons of the \mtb distribution between the predicted background and observed data for both analyses are shown in Figs.~\ref{fig:lep_mtb} and~\ref{fig:had_mtb}.
We observe good agreement between the predicted SM background processes and the observed data,
and proceed to set upper limits at 95\% CL on the $\PWRpr$ boson production cross section for masses between 1 and 3\TeV.
Limits on the cross section of $\PWRpr$ boson production are calculated using a Bayesian method with a flat signal prior, using the \textsc{theta} package~\cite{theta-stat}.
The Bayesian approach uses a binned likelihood to calculate 95\% CL upper limits on the product of the signal production cross section and the branching fraction $\sigma(\Pp\Pp\to\PWRpr)\,\mathcal{B}(\PWRpr\to\PQt\PQb)$.
The computation takes into account all systematic uncertainties given in Section~\ref{sec:syst}, as well as statistical uncertainties related to the backgrounds,
which are incorporated using the "Barlow--Beeston lite" method~\cite{barlowbeeston,Conway:2011in}.
All rate uncertainties are included as nuisance parameters with log-normal priors.

The leptonic analysis separates events into four independent categories according to the lepton type (electron or muon) and the number of b-tagged jets in the first two leading \pt jets (1 or 2).
This improves the sensitivity of the analysis.
In the leptonic analysis, the \mtb distribution is binned to reduce uncertainties from the number of events in each sample.
The binning is as follows:
9 bins with widths of
200\GeV from 400 to 2200\GeV, 1 bin of width 400\GeV from 2200 to 2600\GeV, and 1 bin for 2600\GeV and above.
In the hadronic analysis, the \mtb distribution is binned using 50\GeV bins from 0 to 2100\GeV,
100\GeV bins from 2100 to 2500\GeV, and 1 bin for 2500\GeV and above.

\begin{figure}[htb]
 \centering
  \includegraphics[width=0.48\textwidth]{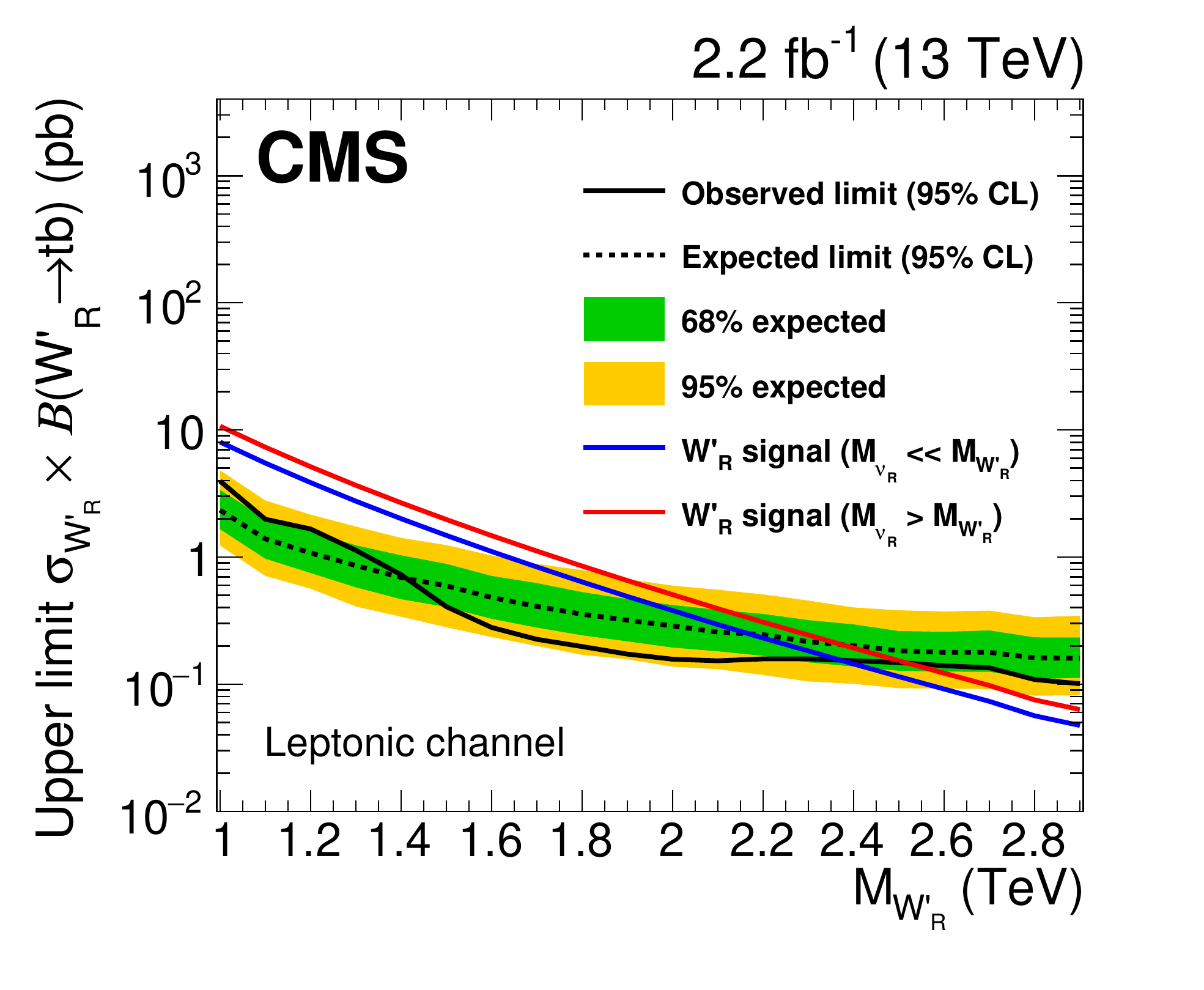}
  \includegraphics[width=0.48\textwidth]{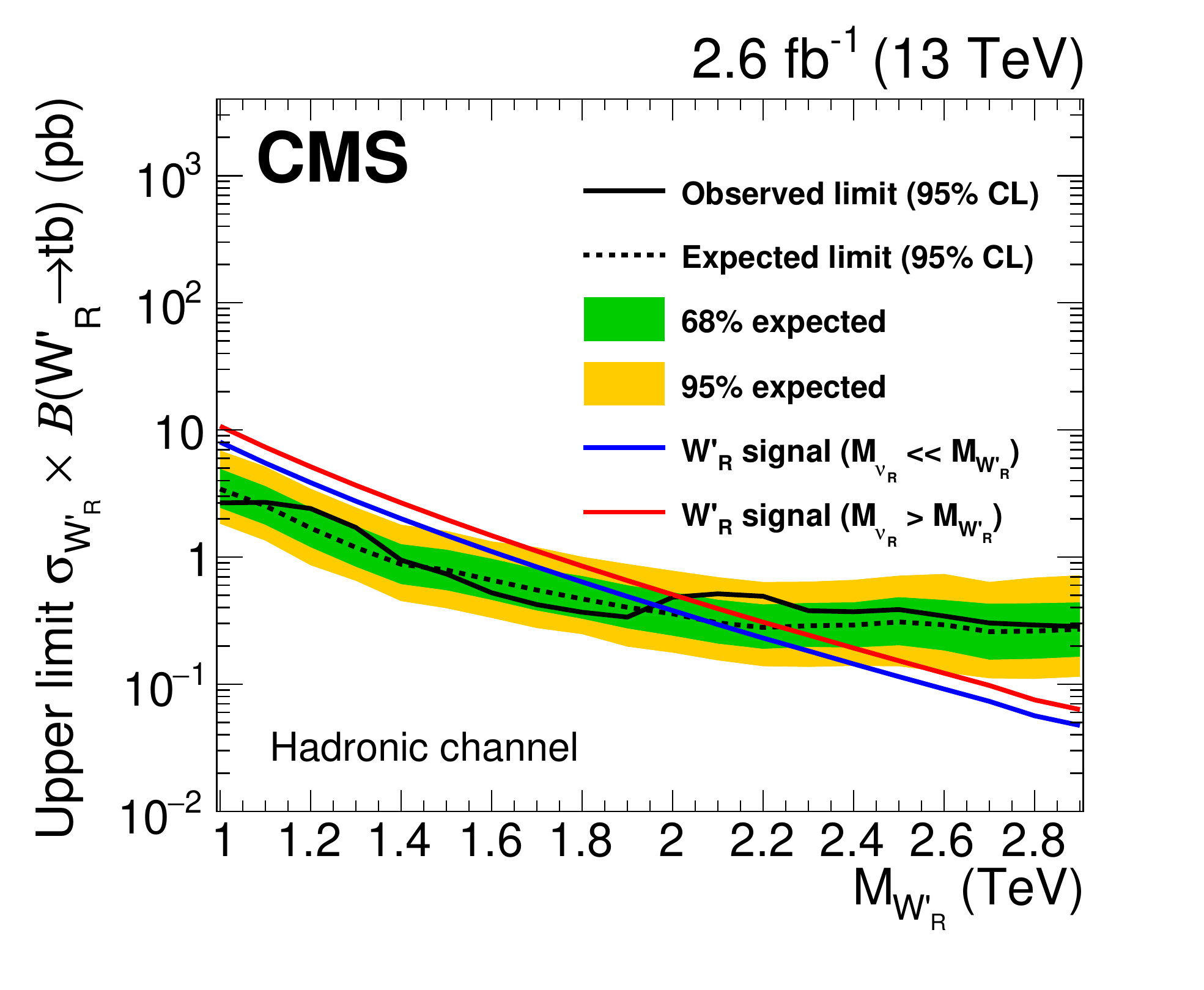}
 \caption{The 95\% CL upper limit on the $\PWRpr$ boson production cross section, separately for the leptonic (left) and hadronic (right) analyses. Masses for which the theoretical cross section is above the observed upper limit are excluded at 95\% CL.\label{fig:sepxseclim}}
\end{figure}

Results from the two analyses are shown separately in Fig.~\ref{fig:sepxseclim}.
The leptonic and hadronic analyses are able to exclude $\PWRpr$ boson masses below 2.4 and 2.0\TeV, respectively.

In combining the two analyses, a joint likelihood is used to simultaneously consider all categories.
We treat the uncertainties related to jet energy scale and resolution, luminosity, pileup, b tagging scale factors, and PDF as fully correlated.
All other uncertainties are considered to be uncorrelated.

\begin{figure}[htb]
 \centering
 \includegraphics[width=0.7\textwidth]{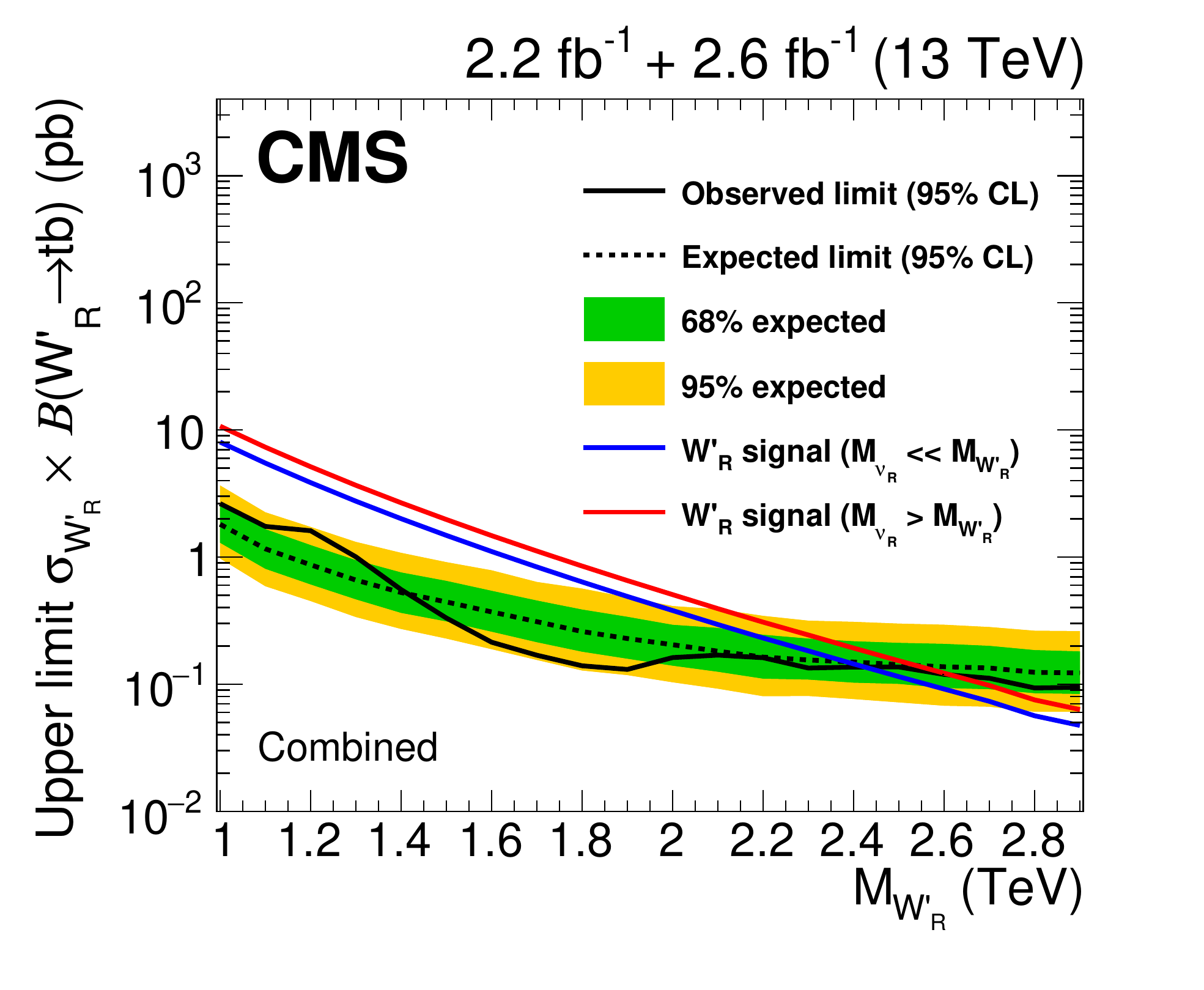}
 \caption{The 95\% CL upper limit on the $\PWRpr$ boson production cross section for the combined leptonic and hadronic analyses. Masses for which the theoretical cross section is above the observed upper limit are excluded at 95\% CL.\label{fig:combxseclim}}
\end{figure}

The combined upper limit on $\PWRpr$ boson production cross section at 95\% CL is shown in Fig.~\ref{fig:combxseclim}.
The observed and expected 95\% CL upper limits are 2.5 and 2.4\TeV, respectively.
This represents a significant improvement over the results from the individual analyses.

\section{Summary\label{sec:summary}}

Searches have been reported for a heavy $\PWRpr$ boson resonance decaying into a top and a bottom quark in data from proton-proton collisions at $\sqrt{s}=13$\TeV collected with the CMS detector.
Analysis of the leptonic and hadronic channels is based on an integrated luminosity of 2.2 and 2.6\fbinv, respectively.
No evidence is observed for the production of a $\PWRpr$ boson, and upper limits at 95\% confidence level on $\sigma(\Pp\Pp\to\PWRpr)\,\mathcal{B}(\PWRpr\to\PQt\PQb)$ are determined as a function of the $\PWRpr$ boson mass.
After combining the two analyses, the upper limits at 95\% confidence level are compared to the predicted $\PWRpr$ boson production cross sections. $\PWRpr$ bosons are excluded for masses less than 2.4\TeV if $M_{\PWRpr}\gg M_{\nu_\mathrm{R}}$, and less than 2.6\TeV if $M_{\PWRpr}<M_{\nu_\mathrm{R}}$.
These results represents the most stringent limits published in the tb decay channel.
\begin{acknowledgments}
We congratulate our colleagues in the CERN accelerator departments for the excellent performance of the LHC and thank the technical and administrative staffs at CERN and at other CMS institutes for their contributions to the success of the CMS effort. In addition, we gratefully acknowledge the computing centers and personnel of the Worldwide LHC Computing Grid for delivering so effectively the computing infrastructure essential to our analyses. Finally, we acknowledge the enduring support for the construction and operation of the LHC and the CMS detector provided by the following funding agencies: BMWFW and FWF (Austria); FNRS and FWO (Belgium); CNPq, CAPES, FAPERJ, and FAPESP (Brazil); MES (Bulgaria); CERN; CAS, MoST, and NSFC (China); COLCIENCIAS (Colombia); MSES and CSF (Croatia); RPF (Cyprus); SENESCYT (Ecuador); MoER, ERC IUT, and ERDF (Estonia); Academy of Finland, MEC, and HIP (Finland); CEA and CNRS/IN2P3 (France); BMBF, DFG, and HGF (Germany); GSRT (Greece); OTKA and NIH (Hungary); DAE and DST (India); IPM (Iran); SFI (Ireland); INFN (Italy); MSIP and NRF (Republic of Korea); LAS (Lithuania); MOE and UM (Malaysia); BUAP, CINVESTAV, CONACYT, LNS, SEP, and UASLP-FAI (Mexico); MBIE (New Zealand); PAEC (Pakistan); MSHE and NSC (Poland); FCT (Portugal); JINR (Dubna); MON, RosAtom, RAS, RFBR and RAEP (Russia); MESTD (Serbia); SEIDI, CPAN, PCTI and FEDER (Spain); Swiss Funding Agencies (Switzerland); MST (Taipei); ThEPCenter, IPST, STAR, and NSTDA (Thailand); TUBITAK and TAEK (Turkey); NASU and SFFR (Ukraine); STFC (United Kingdom); DOE and NSF (USA).

\hyphenation{Rachada-pisek} Individuals have received support from the Marie-Curie program and the European Research Council and Horizon 2020 Grant, contract No. 675440 (European Union); the Leventis Foundation; the A. P. Sloan Foundation; the Alexander von Humboldt Foundation; the Belgian Federal Science Policy Office; the Fonds pour la Formation \`a la Recherche dans l'Industrie et dans l'Agriculture (FRIA-Belgium); the Agentschap voor Innovatie door Wetenschap en Technologie (IWT-Belgium); the Ministry of Education, Youth and Sports (MEYS) of the Czech Republic; the Council of Science and Industrial Research, India; the HOMING PLUS program of the Foundation for Polish Science, cofinanced from European Union, Regional Development Fund, the Mobility Plus program of the Ministry of Science and Higher Education, the National Science Center (Poland), contracts Harmonia 2014/14/M/ST2/00428, Opus 2014/13/B/ST2/02543, 2014/15/B/ST2/03998, and 2015/19/B/ST2/02861, Sonata-bis 2012/07/E/ST2/01406; the National Priorities Research Program by Qatar National Research Fund; the Programa Clar\'in-COFUND del Principado de Asturias; the Thalis and Aristeia programs cofinanced by EU-ESF and the Greek NSRF; the Rachadapisek Sompot Fund for Postdoctoral Fellowship, Chulalongkorn University and the Chulalongkorn Academic into Its 2nd Century Project Advancement Project (Thailand); and the Welch Foundation, contract C-1845.
\end{acknowledgments}
\clearpage
\bibliography{auto_generated}

\cleardoublepage \appendix\section{The CMS Collaboration \label{app:collab}}\begin{sloppypar}\hyphenpenalty=5000\widowpenalty=500\clubpenalty=5000\textbf{Yerevan Physics Institute,  Yerevan,  Armenia}\\*[0pt]
A.M.~Sirunyan, A.~Tumasyan
\vskip\cmsinstskip
\textbf{Institut f\"{u}r Hochenergiephysik,  Wien,  Austria}\\*[0pt]
W.~Adam, F.~Ambrogi, E.~Asilar, T.~Bergauer, J.~Brandstetter, E.~Brondolin, M.~Dragicevic, J.~Er\"{o}, M.~Flechl, M.~Friedl, R.~Fr\"{u}hwirth\cmsAuthorMark{1}, V.M.~Ghete, J.~Grossmann, N.~H\"{o}rmann, J.~Hrubec, M.~Jeitler\cmsAuthorMark{1}, A.~K\"{o}nig, I.~Kr\"{a}tschmer, D.~Liko, T.~Madlener, T.~Matsushita, I.~Mikulec, E.~Pree, D.~Rabady, N.~Rad, H.~Rohringer, J.~Schieck\cmsAuthorMark{1}, M.~Spanring, D.~Spitzbart, J.~Strauss, W.~Waltenberger, J.~Wittmann, C.-E.~Wulz\cmsAuthorMark{1}, M.~Zarucki
% ...
\vskip\cmsinstskip
1:~~Also at Vienna University of Technology, Vienna, Austria\\
% ...

\end{sloppypar}
\end{document}